# Rubber-to-glass adhesion between a rigid sphere and a shape memory polymer substrate of finite thickness

Changhong Linghu [1,†], Wentao Mao [1,†], Haoyu Jiang [1], Huajian Gao [2,*], K. Jimmy Hsia [1,3,*]

[1] *School of Mechanical and Aerospace Engineering, Nanyang Technological University, 50 Nanyang Avenue, Singapore 639798, Singapore*

[2] *Mechano-X Institute, Applied Mechanics Laboratory, Department of Engineering Mechanics, Tsinghua University, Beijing 100084, China.*

[3] *School of Chemistry, Chemical Engineering and Biotechnology, Nanyang Technological University, 50 Nanyang Avenue, Singapore 639798, Singapore*

† *These authors contributed equally*

\* *To whom the correspondence should be addressed:*

\* *gao.huajian@tsinghua.edu.cn*; *kjhsia@ntu.edu.sg*

**Abstract:** Shape memory polymers (SMPs) are emerging as innovative smart adhesive materials with broad application potential. Compared to conventional elastomeric adhesives, SMP adhesives are distinguished by the so-called rubber-to-glass (R2G) adhesion, which involves contact in the rubbery state followed by detachment in the glassy state. This process, through a shape-locking effect, enhances adhesion strength by more than an order of magnitude compared to conventional adhesive contact. Here, we investigate the fundamental problem of a rigid sphere undergoing R2G adhesion with an SMP substrate of finite thickness through experiments, finite element (FE) simulations, and theoretical modeling. It is demonstrated that during press-in, the contact problem can be modeled as a rigid oblate spheroid contacting an infinite substrate, while the pull-off process can be described by a modified ball-and-socket model. These equivalent models yield practically useful analytical solutions for the contact radius during press-in and the R2G adhesion force during pull-off. A critical thickness-to-contact-radius ratio of around 5 is identified, below which the thickness effect becomes significant. These insights provide valuable guidance for the design and application of SMP-based smart adhesives.

**Highlights:**

- A comprehensive study of the SMP R2G adhesion with adhesives of finite-thickness

- Revealing the underlying mechanisms of substrates' thickness effect on R2G adhesion

- Oblate spheroid equivalence analytical model that predicts the contact radius

- A modified ball-and-socket model that predicts SMP R2G adhesion forces accurately

- Identifying the critical thickness-to-contact-radius ratio for finite thickness effect

# 1. Introduction

Strong and on-demand adhesives (Jin et al., 2016; Wu et al., 2023; Yin et al., 2022) are crucial in many applications, including robotics (Gu et al., 2018; Jiang et al., 2017; Levine et al., 2022; Li et al., 2022a; Li et al., 2022b; Linghu et al., 2023a; Luo et al., 2022; Ruotolo et al., 2021), wearables (Gao et al., 2019; Li et al., 2019; Liu et al., 2023; Shin et al., 2024; Zhang et al., 2019), advanced manufacturing (Dai et al., 2019; Huang et al., 2019; Jian et al., 2024; Jiang et al., 2024; Linghu et al., 2019; Linghu et al., 2018; Wu et al., 2021; Zhang et al., 2021a; Zhang et al., 2021c), and biomedical engineering (Deng et al., 2021; Wang et al., 2022; Wang et al., 2015; Yuk et al., 2021). Dry adhesives (Arzt et al., 2021; Hensel et al., 2018), which rely on van der Waals forces, are clean, sustainable, energy-efficient, and reusable, attracting increasing research community attention. Over the past few decades, researchers have used these smart dry adhesives to develop devices with fibrillar microstructures mimicking those found in geckos (Autumn et al., 2000; Chen and Gao, 2007; Gao et al., 2005; Gao and Yao, 2004) to realize strong and reversible adhesion on various surfaces. However, gecko-like adhesives are still limited to an adhesion strength of around 100 kPa (Labonte et al., 2016; Linghu et al., 2023b; Linghu et al., 2024), which is insufficient in applications requiring heavy load capabilities exceeding tens of kilograms (Tao et al., 2023; Yang et al., 2023).

Recently, innovative smart adhesives using shape memory polymers (SMPs) (Luo et al., 2023; Xia et al., 2021; Zhao et al., 2023; Zheng et al., 2015), inspired by the stiffening and locking mechanism of snail mucus on surfaces (Cho et al., 2019; Linghu et al., 2023b; Shi et al., 2024), have gained popularity due to their significantly enhanced adhesion strength (Linghu et al., 2023b; Linghu et al., 2023c; Linghu et al., 2020; Zhang et al., 2024; Zhao et al., 2021) compared to conventional smart adhesives based on soft elastomers. SMPs exhibit tunable modulus upon stimulation, forming conformal contact with rough surfaces in their soft rubbery phase, and then locking the contact upon transitioning to the stiff glassy phase. The adhesion method — utilizing SMPs to contact in the rubbery phase and attach in the glassy phase through rubber-to-glass (R2G) phase transitions — enhances adhesion by more than an order of magnitude, offering superior adaptability to a variety of rough surfaces (Linghu et al., 2023b; Linghu et al., 2023c). Consequently, SMP adhesives have greatly broadened the applications of smart dry adhesives, advancing multiple applications such as universal grippers (Linghu et al., 2020; Luo et al., 2022; Son et al., 2023; Son and Kim, 2020; Ze et al., 2020; Zhao et al., 2021; Zhuo et al., 2020), versatile adhesive hooks (Eisenhaure and Kim, 2018; Linghu et al., 2024; Son and Kim, 2021), micro-nano transfer printing devices (Eisenhaure and Kim, 2016; Kim et al., 2024; Linghu et al., 2020; Luo et al., 2021; Luo et al., 2024; Xue et al., 2015; Zhang et al., 2021b), and wearable electronics (Wang et al., 2021; Yi et al., 2019; Zhang et al., 2019).

One key feature that distinguishes the adhesive behaviours of SMP adhesives from conventional elastomeric adhesives is the tremendous adhesion enhancement enabled by the R2G transition. Understanding and predicting SMP R2G adhesion are critical for their design and application. For a while, some researchers have attributed the enhancement of SMP R2G adhesion to the modulus effect (Eisenhaure and Kim, 2014; Eisenhaure et al., 2013; Tan et al., 2019). But recently, Linghu et al (Linghu et al., 2023c) have shown, through studying the SMP R2G adhesion between a rigid sphere and an SMP substrate (Figs. 1-i to v), that the shape-locking effect, rather than the modulus effect, is the primary factor that is responsible for the tremendous enhancement in SMP R2G adhesion. As illustrated in Fig. 1, the R2G adhesion process can be divided into three stages: press-in, phase transition, and pull-off. During press-in, a rigid sphere indents into the SMP substrate in its rubbery state. Once the desired indentation depth is reached, the SMP is brought to transition into its glassy state, causing the deformation induced in the rubbery state to be 'frozen' due to the shape-locking effect. The adhesion under these conditions generates a pull-off force much greater than that associated with conventional adhesive contact between elastic bodies. Linghu et al. (Linghu et al., 2023c) demonstrated that the effect of conformal contact 'locking' in SMP R2G adhesion could be approximately modelled as a flat punch (FPA) on an elastomeric substrate (Hui et al., 2004; Jiang et al., 2014), which leads to an analytical solution for the SMP R2G pull-off force.

While the FPA shows good agreement with experimental results, substantial deviations arose in cases involving a large sphere radius ($R = 50$ mm) and deeper indentations (Linghu et al., 2023c) due to the finite-thickness effect of the substrate (Bartlett and Crosby, 2013; Hensel et al., 2019; Müller and Müser, 2023; Peng et al., 2020; Webber et al., 2003). In many practical applications, SMP adhesives typically have a finite thickness (Figs. 1-vi to x) due to considerations such as ease of fabrication (Linghu et al., 2020), user comfort (Yi et al., 2019), and actuation speed (Son et al., 2023; Ze et al., 2020). However, to date, the finite-thickness effect on SMP R2G adhesion and the SMP R2G pull-off forces under finite-thickness conditions have not been studied.

In this paper, we investigate the finite thickness effects of SMP substrate on R2G adhesion. We develop an analytical model and perform finite element (FE) simulations that accurately predict the SMP R2G pull-off force for substrates of varying thicknesses. The rest of the paper is organized as follows. Section 2 introduces an oblate spheroid equivalence (OSE) model to describe the contact radius during the press-in process, and a modified ball-and-socket (MBS) model to evaluate the adhesion force during pull-off for the fundamental problem of a rigid sphere contacting with SMP substrates of finite thickness. These models establish scaling laws for the contact radius and pull-off forces as functions of the substrate-thickness-to-contact-radius ratio. Section 3 presents FE simulations conducted across a series of sphere radii and substrate thickness values. In Section 4, empirical expressions for the contact radius and SMP R2G pull-off forces are obtained by fitting their scaling laws to the FE simulation results. These expressions are validated against an additional set of FE simulation results and experimental results from previous research (Linghu et al., 2023c). Predictions from FPA models under finite-thickness conditions are also compared and discussed. Conclusions are presented in Section 5.

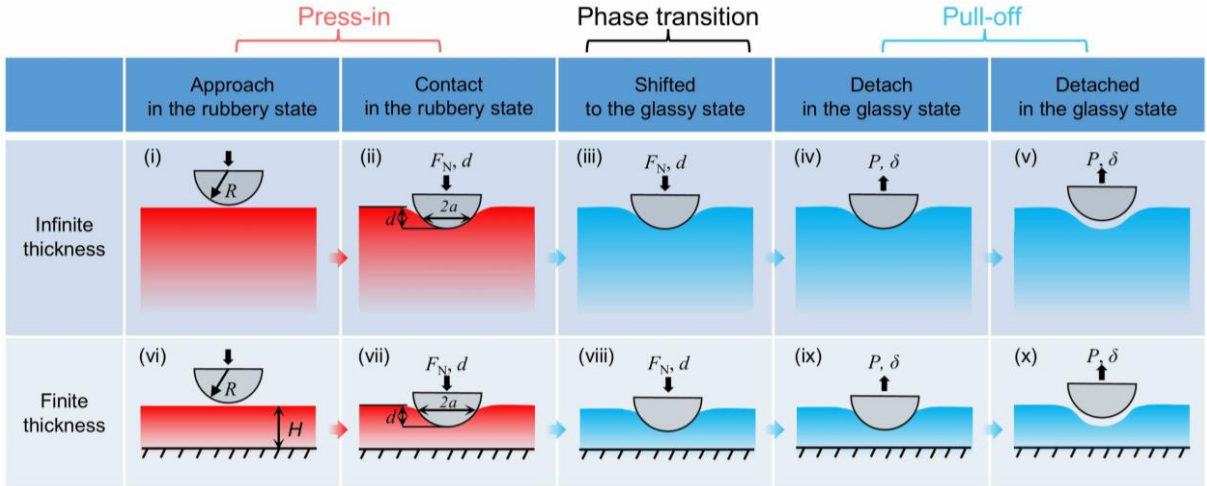

**Fig. 1. Schematic illustrations of rubber-to-glass (R2G) adhesion between a rigid sphere and an SMP substrate with (i-v) infinite thickness and (vi-x) finite thickness**. Initially (**i-ii; vi-vii**), the rigid sphere is pressed into the SMP substrate in its soft, rubbery state. The substrate then undergoes the R2G phase transition (**iii; viii**), shifting to the glassy state and locking the contact configuration. During pull-off (**iv-v; ix-x**), the rigid sphere is detached from the now glassy substrate, resulting in ultra-strong R2G adhesion.

## 2.  Theoretical modeling

### 2.1.  Modeling of the press-in process

As reported in the literature (Linghu et al., 2023c; Linghu et al., 2020), SMP R2G adhesion is strongly dependent on the indentation depth $d$, in contrast to the adhesive contact behavior of elastic bodies where adhesion forces are independent of the indentation depth. This dependence of SMP R2G adhesion on indentation depth arises from the shape-locking effect and the resulting 'freezing' of the contact radius $a$. Consequently, it is crucial to first establish the relationship between the contact radius and the indentation depth, i.e., the *d-a* relations.

Neglecting the influence of adhesion during press-in, the *d-a* relation can be obtained from the Hertz model (Hertz, 1882) as $a = \sqrt{Rd}$ for small deformations at shallow indentations, and from the modified Hertz model (Guo et al., 2020) as $a = \sqrt{Rd - d^2/4}$ for large deformations at deep indentations, both assuming an infinitely thick substrate.

When the substrate-thickness-to-contact-radius ratio is moderate, the assumptions of the Hertz and modified Hertz models are no longer valid, thus the *d-a* relation depends on substrate thickness. Yu et al. (Yu et al., 1990) analyzed the contact of a rigid sphere on a thin-film substrate of linear elastic materials and obtained a numerical solution for the *d-a* relation across varying substrate thicknesses. However, this approach does not provide a single analytical expression for the *d-a* relation across different thicknesses.

To derive a single analytical expression for the *d-a* relation of a rigid sphere on a substrate of finite thickness, we identify an equivalence between spherical contact on a finite-thickness substrate (Fig. 2a) and oblate spheroid contact on an infinite-thickness substrate (Fig. 2b).

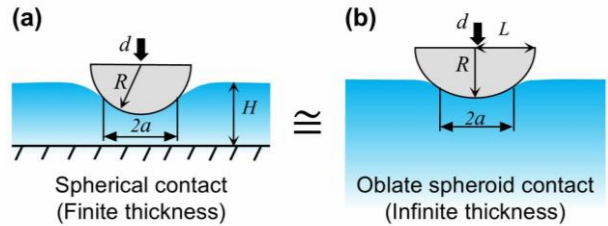

**Fig. 2. The equivalence between a rigid sphere in contact with an elastic substrate of finite thickness and a rigid oblate spheroid in contact with an elastic substrate of infinite thickness (during press-in).** (**a**) The original problem and (**b**) the equivalent problem.

The *d-a* relation for the contact of oblate spheroid indenters on an infinite-thickness substrate is given by (Popov et al., 2019)

$$d = a \cdot \left( \frac{R}{L} \right) \cdot \operatorname{arctanh}\left( \frac{a}{R} \cdot \frac{R}{L} \right), \qquad (1)$$

where *L* and *R* are the major and minor axes of the oblate spheroid, respectively. When *R/L*=1, the oblate spheroid becomes a sphere.

From Eq. (1), an equivalence can be established between the spherical and oblate spheroid contact problems. The numerical results of Yu et al. (Yu et al., 1990) indicate that the relation between the contact radius *a* and the indentation depth *d* is influenced by the relative thickness *H/a*. In fact, a decrease in the relative substrate thickness *H/a* leads to an increase in the contact radius *a* under the same indentation depth in the sphere contact system. This effect is equivalent to an increase in the major axis *L* in the oblate spheroid contact problem under infinite-thickness conditions. Based on this argument, the special contact problem under finite-thickness conditions can be considered analogous to the oblate spheroid contact problem under infinite-thickness conditions. In the equivalent oblate spheroid contact system (Fig. 2b), the minor axis *R* is equal to the radius of the sphere, and the ratio *R/L* of the oblate spheroid indenter can be approximated by an exponential function (as will be shown by the FE simulation results in Section 4) of the thickness-to-contact-radius ratio, *H/a*, as follows:

$$R/L = 1 - \alpha \, \mathrm{e}^{(-\beta \cdot (H/a))}, \qquad (2)$$

where $\alpha$ and $\beta$ are constants to be determined.

Substituting Eq. (2) into Eq. (1), the scaling law for the *d-a* relation of spherical contact problem under finite-thickness conditions can be obtained as

$$d = a[1 - \alpha \, \mathrm{e}^{(-\beta \cdot (H/a))}] \cdot \operatorname{arctanh}([1 - \alpha \, \mathrm{e}^{(-\beta \cdot (H/a))}] \cdot a / R) , \qquad (3)$$

where the constants $\alpha$ and $\beta$ can be determined by fitting the FE simulation results. The fitting results will be presented in the Results section below.

## 2.2. Modeling of the pull-off process

**The pull-off force**

The compliance method (Levine et al., 2022; Linghu et al., 2023a) is employed to derive an expression for the R2G pull-off force of the contact system. In this process, the deformation in the SMP adhesive substrate, incurred during the press-in stage, is locked by the shape-locking effect of the SMP, as illustrated in Fig. 1.

The total energy of the system $U_{tot}$ during pull-off can be divided into three components: surface energy $U_s$, strain energy of the substrate $U_e$, and potential energy of external force $U_p$ as

$$U_{tot} = U_s + U_e + U_p. \tag{4}$$

The surface energy $U_s$ can be calculated by multiplying the work of adhesion $w$ with the actual contact area, as

$$U_s = -A_{real}w. \tag{5}$$

where the actual contact area can be calculated according to

$$A_{real} = \int_0^\theta 2\pi R \sin(\varphi) R d\varphi. \tag{6}$$

With the integration limit, $\theta$, given by

$$\cos\theta = \frac{\sqrt{R^2 - a^2}}{R}, \tag{7}$$

the contact area is obtained as

$$A_{real} = 2\pi R^2 \left(1 - \frac{\sqrt{R^2 - a^2}}{R}\right). \tag{8}$$

The strain energy of the substrate $U_e$ during the pull-off is given by

$$U_e = \frac{1}{2}P\delta = \frac{1}{2}\frac{P^2}{K}, \tag{9}$$

where $\delta$ is the pull-off displacement and $K$ is the stiffness of the deformed substrate after the R2G transition, with the deformation locked in the glassy state.

The potential energy of the external force $U_p$ is

$$U_p = -P\delta = -\frac{P^2}{K} \tag{10}$$

From Eqs. (5), (9), and (10), the total energy is obtained as follows:

$$U_{\text{tot}} = -2\pi R^2 \left(1 - \frac{\sqrt{R^2 - a^2}}{R}\right) w - \frac{P^2}{2K} . \tag{11}$$

Assuming the detachment is energy-controlled (Linghu et al., 2023a), the total energy reaches a minimum at the moment of detachment, resulting in the critical pull-off condition, as

$$\frac{\partial U_{tot}}{\partial a} = -\frac{2\pi Raw}{\sqrt{R^2 - a^2}} + \frac{P^2 \partial K / \partial a}{2K^2} = 0 . \tag{12}$$

Therefore, the final expression for the pull-off force is

$$P_{\text{c}} = \sqrt{\frac{4\pi RawK^2}{\partial K / \partial a \sqrt{R^2 - a^2}}} , \tag{13}$$

which relates the pull-off force to the stiffness $K$ of the adhesion system.

**The pull-off stiffness**

According to Eq. (13), the key to obtaining the pull-off force lies in determining the stiffness of the adhesive system at the pull-off point. The conformal contact formed during press-in in the rubbery state is locked in the glassy state after the R2G transition. Under finite-thickness conditions, both the substrate thickness and the local curvature of the contact surface significantly influence the system's stiffness. In this section, we discuss various methods to obtain analytical, approximate solutions for the stiffness of the adhesion system under finite-thickness conditions.

<u>Generalization of the FPA model</u>

Given that the conformal contact is locked after the R2G transition, Linghu et al. (Linghu et al., 2023c) employed the FPA model to predict SMP R2G adhesion of a rigid sphere with an infinite-thickness SMP adhesive substrate. This FPA model can be extended to account for the finite-thickness effects on the R2G adhesion problem.

Several flat punch models can currently account for thickness effects (see Appendix A for details). Kendall et al. (Kendall, 1971) proposed a model for the adhesion force of a flat punch on thin films. Yang (Yang, 2006) introduced a modified stiffness expression based on linear elastic mechanics analysis. Shull et al. (Shull et al., 1998) derived an empirical formula for the pull-off force, considering thickness effects on stiffness through an analysis of FE simulation results. Peng et al. (Peng et al., 2020) provided an alternative expression applicable to a broad range of thicknesses through theoretical analysis. The expressions for stiffness and pull-off forces derived from these models are summarized in **Table I**.

**Table I.** Summary of existing flat-punch models considering the substrate's thickness effect

| Substrate condition | Stiffness $K$ | Pull-off force $P_c$ | Reference |
|---|---|---|---|
| Thin film | $K_{\text{Kendall}} = \dfrac{\pi a^2 E}{3(1-2v)H}$ | $P_{c,\,\text{Kendall}} = \pi a^2 \sqrt{\dfrac{2Ew}{3(1-2v)H}}$ | Kendall, 1971 |
| | $K_{\text{Yang}} = \dfrac{(1-v)\pi a^2 E}{(1+v)(1-2v)H}$ | $P_{c,\,\text{Yang}} = \pi a^2 \sqrt{\dfrac{2Ew(1-v)}{(1+v)(1-2v)H}}$ | Yang, 2006 |
| Finite thickness | $K_{\text{Shull}} = \dfrac{2Ea}{1-v^2}\left[1 + \left(\dfrac{0.75}{\left((a/h)+(a/h)^3\right)} + \dfrac{2.8(1-2v)}{(a/h)}\right)^{-1}\right]$ | $P_{c,\,\text{Shull}} = \sqrt{\dfrac{4\pi Raw K_{\text{Shull}}^{\,2}}{\partial K_{\text{Shull}}/\partial a}}$ | Shull et al., 1998 |
| | $K_{\text{Peng}} = \dfrac{2Ea}{1-v^2}k(H/a, v)$ | $P_{c,\,\text{Peng}} = \sqrt{\dfrac{8\pi w Ea^3}{1-v^2}\psi\left(\dfrac{H}{a}, v\right)}$ | B. Peng et al., 2020 |

Modified ball-and-socket (MBS) model

Directly generalizing FPA models to finite-thickness conditions, without accounting for the impact of substrate curvature, leads to large errors at greater indentation depths. This limitation was demonstrated by Linghu et al. (Linghu et al., 2023c) using the model by Peng et al. (Peng et al., 2020). To accurately predict the SMP R2G pull-off force under finite-thickness conditions, it is necessary to account for both the thickness and the substrate curvature effects. The configuration of a rigid sphere locked within the SMP substrate after the R2G transition closely resembles that of the ball-and-socket contact model (Heß and Forsbach, 2021), as illustrated in Fig. 3a, which is often used to model bone joint systems.

The Hertzian contact model assumes small deformation and the substrate in its undeformed configuration. This approach overlooks the curvature effect caused by substrate deformation. In contrast, the ball-and-socket model considers the undeformed substrate with a curvature (Figs 3a, b). The mechanics of the ball-and-socket model is similar to that of the SMP R2G adhesion upon shape-locking, where the substrate shares the same local curvature as the spherical indenter (Fig. 3c). Based on this analogy, an equivalent ball-and-socket model (Fig. 3d) is proposed for SMP R2G adhesion.

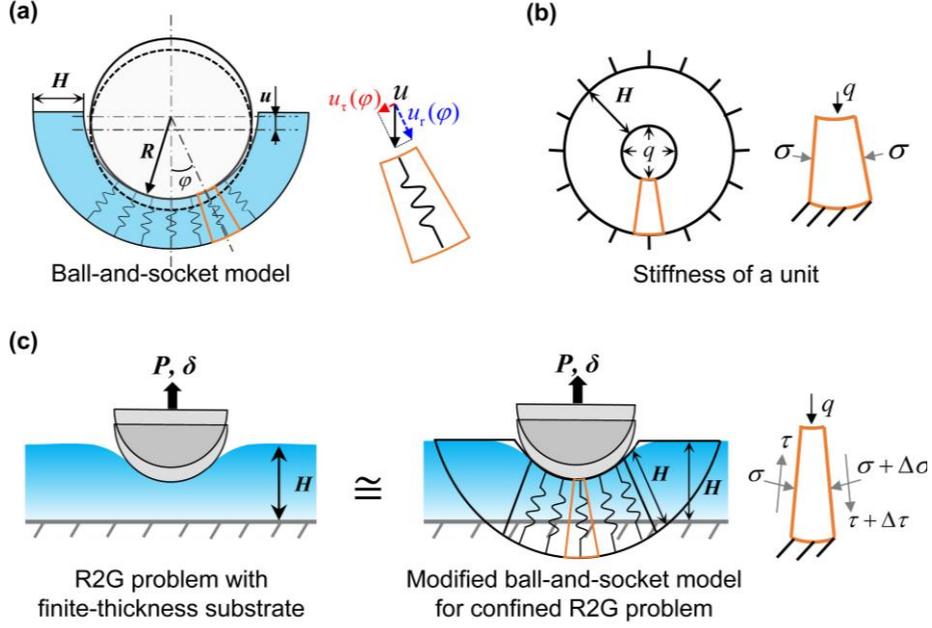

**Fig. 3. A modified ball-and-socket (MBS) model of R2G adhesion between a rigid sphere and an SMP substrate of finite thickness.** (**a**) A ball-and-socket model under compressive displacement $u$, with a magnified view showing the displacement components of a single unit at angle $\varphi$. (**b**) Stiffness calculation of a single unit in the ball-and-socket model based on the thick-walled spherical vessel problem, where the outer surface is fixed, and the inner surface is subjected to uniform pressure. The magnified view illustrates the stress state of a single unit in the thick-walled spherical vessel. (**c**) The original model of a rigid sphere being pulled off from an SMP substrate of finite thickness and its equivalent ball-and-socket model with a magnified view depicting the stress state of a single unit.

In the ball-and-socket model, the substrate is considered as a collection of independent springs, each making normal contact with the contact surface (Heß and Forsbach, 2021). The deformation of a single element is shown in a zoom-in view in Fig. 3a. The compressive displacement perpendicular to the contact surface at a point with an angle $\varphi$ can be expressed in terms of the vertical displacement $u$ of the spherical indenter, as

$$u_r(\varphi) = u\cos(\varphi).$$ (14)

The stiffness of an individual element can be derived from the problem of a thick-walled spherical vessel, fixed at the outer surface and subjected to internal pressure (Fig. 3b) (Roark and Young, 2002), as

$$k = \frac{2(1-\nu)+(1+\nu)\alpha^3}{(1+\nu)(1-2\nu)(1-\alpha^3)}\frac{E}{R},$$ (15)

$$\alpha = \frac{R}{R+H}.$$ (16)

The radial stress distribution is then given by

$$p_{\mathrm{r}}(\varphi) = k u_{\mathrm{r}}(\varphi). \tag{17}$$

By integrating the stress along the vertical direction of all the individual elements in the ball-and-socket model, the total force in the vertical direction can be derived as,

$$P_y = 2\pi R^2 \int_0^\theta p_{\mathrm{r}}(\varphi) \cos(\varphi) \sin(\varphi) d\varphi = \frac{2\pi R^2 k u}{3} [1 - (1 - (\frac{a}{R})^2)^{\frac{3}{2}}]. \tag{18}$$

Thus, the contact stiffness of the ball-and-socket model can be obtained,

$$K = \frac{\partial P_y}{\partial u} = \frac{2\pi R^2 k}{3} [1 - (1 - (\frac{a}{R})^2)^{\frac{3}{2}}]. \tag{19}$$

For a thick-walled spherical vessel subjected to internal pressure, there is no shear stress on the socket wall due to symmetry. However, in the equivalent ball-and-socket model of the SMP R2G adhesion (Fig. 3c), where the thickness of the socket is assumed to be equal to the undeformed SMP substrate thickness $H$, gradient exists in both the normal and shear stress components on each element of the side wall, dropping to zero at the top surface of the substrate. To account for this, a correction function $\varsigma(a/H)$ is introduced, yielding the stiffness expression for the MBS model as follows:

$$K_{\mathrm{m}} = \varsigma\left(\frac{a}{H}\right) \cdot \frac{2\pi R^2 k}{3} [1 - (1 - (\frac{a}{R})^2)^{\frac{3}{2}}]. \tag{20}$$

Substituting Eq. (20) into the Eq. (13) yields the expression for the pull-off force as

$$P_{\mathrm{c}} = \eta(\frac{a}{H}) \cdot \bar{P}_{\mathrm{c}}, \tag{21}$$

where

$$\bar{P}_{\mathrm{c}} = \sqrt{\frac{8wk\pi^2 R^6 [1 - (1 - (\frac{a}{R})^2)^{\frac{3}{2}}]^2}{9(R^2 - a^2)}}, \tag{22}$$

and

$$\eta(a/H) = P_{\mathrm{c}} / \bar{P}_{\mathrm{c}} \tag{23}$$

is the correction function of the pull-off force to be determined.

## 3. FE simulation setup

To determine the specific functional forms for the contact radius in Eq. (3) from the oblate spheroid equivalence (OSE) model and the R2G pull-off force in Eq. (21) from the MBS model, FE simulations of the R2G adhesion of a rigid sphere on SMP substrates with varying thicknesses and sphere radii were conducted using ABAQUS/Standard (Dassault Systèmes). The simulations employ constitutive laws of an epoxy SMP adhesive (Eisenhaure and Kim, 2016; Eisenhaure et al., 2016; Linghu et al., 2023b; Linghu et al., 2023c; Zheng et al., 2015) that have been extensively studied. This SMP exhibits a modulus of approximately 1 MPa in the rubbery state and 1 GPa in the glassy state, along with ultra-strong R2G adhesion in the MPa range.

An axisymmetric model, shown in Fig. 4, was considered, with the substrate fixed at the bottom. The radius of the substrate was three times that of the sphere. The simulations utilized a thermomechanical constitutive model for the SMP (Liu et al., 2020; Tobushi et al., 1997; Tobushi et al., 2001).

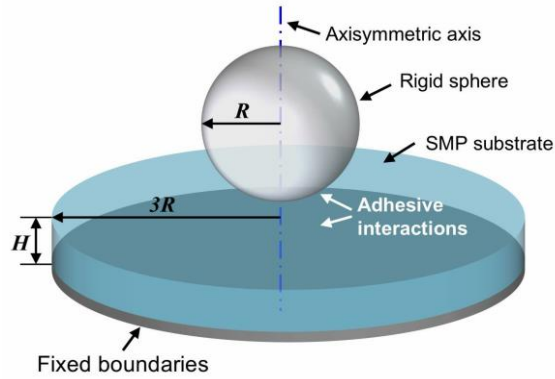

**FIG. 4. Finite element simulation of a rigid sphere in R2G adhesion with an SMP substrate with finite thickness $H$.**

RAX2 (rigid, axisymmetric 2-node) elements were used for the rigid sphere, while CAX4 (continuum, axisymmetric 4-node) elements were applied to the SMP substrate. Adhesive interactions were established during the contact process and were modeled using a linear traction-separation law. A quadratic nominal stress criterion was employed for initiating damage, with damage evolution governed by a power-law ($\alpha$=2) mixed-mode behavior (Camanho et al., 2003; Sancaktar, 1995). The adhesive parameters between the sphere and the SMP substrate were set as follows: $\sigma_{th} = 1.893 \, \text{MPa}$, $\tau_{th} = 5 \, \text{MPa}$, $G_{IC} = 1.61 \, \text{J/m}^2$ and $G_{IIC} = 3.22 \, \text{J/m}^2$, where $\sigma_{th}$ and $\tau_{th}$ represent the theoretical adhesion strengths in the normal and shear directions, respectively, $G_{\text{IC}}$ and $G_{\text{IIC}}$ denote the critical energy release rates (or work of adhesion) for Mode-I and Mode-II cracks, respectively. More details can be found in (Linghu et al., 2023c).

The simulation results are used to determine the expressions for the contact radius during press-in and the pull-off force during detachment, as well as to validate these expressions across various ranges of sphere radius and substrate thickness.

## 4. Results and discussion

### 4.1. Mechanisms influencing the adhesion

Figures 5a and 5b compare the FE simulation results of the normal (S22) stress distribution in the sphere contact system with an SMP substrate of infinite thickness ($H$=40 mm, $H/a$ > 5) and finite thickness ($H$=3 mm) at the pull-off point, respectively. The radius of the indenter ($R$=20 mm) and the indentation depth ($d$=2 mm) in these simulations are the same. Figure 5c shows the control case of a flat-punch system with a contact radius (8.35 mm) equal to that of the indentation with a spherical indenter under finite-thickness conditions ($R$=20 mm, $H$=3 mm, $d$=2 mm). Figure 5d shows the change in the S22 distribution at the contact interface in the spherical indentation system as the substrate thickness decreases, while Fig. 5e compares the S22 and S12 distributions at the contact interface of the spherical indentation system ($R$=20 mm, $H$=3 mm, $d$=2 mm, Fig. 5b) with that of the flat-punch system of the same contact radius ($a$ =8.35 mm Fig. 5c).

Figure 5d shows that, as the substrate thickness $H$ decreases from 40 mm to 3 mm, the contact radius gradually increases from 6.32 mm (Fig. 5a) to 8.35 mm (Fig. 5b), by 32%. This significant increase in contact radius, due to reduced substrate thickness, directly leads to an increase in the R2G pull-off force during detachment. Additionally, the normal stress S22, which is directly responsible for the pull-off force, becomes more uniform as the substrate thickness decreases. This is primarily attributed to the increased system stiffness of the substrate with decreasing thickness, a phenomenon extensively discussed in the flat-punch system (Hayes et al., 1972; Kendall, 1971; Peng et al., 2020; Shull et al., 1998; Yang, 2006) under finite-thickness conditions.

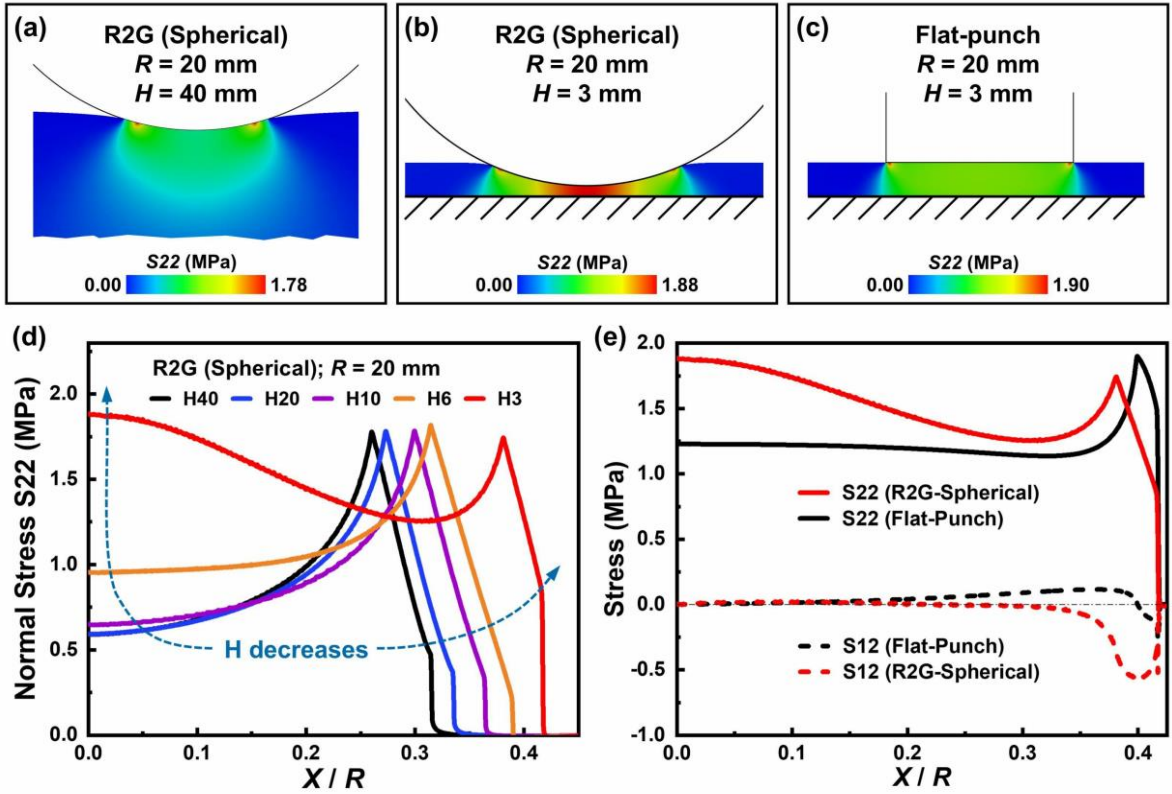

**Fig. 5. FE simulation results showing the deformed configurations and stress distributions at the pull-off point.** (**a–b**) The deformed configuration and normal stress ($S22$) distribution cloud map for a sphere contact system with an SMP substrate of (**a**) infinite thickness ($H$=40 mm) and (b) finite thickness ($H$=3 mm) at the pull-off point. Here, $R$=20 mm and $d$=2 mm. (**c**) The deformed configuration and normal stress ($S22$) cloud map for a flat-punch system, where the contact radius (8.35 mm) is equal to that of the system in (**b**). (**d**) The variation in $S22$ distribution at the contact interface as substrate thickness decreases in the sphere contact system. (**e**) A comparison of the $S22$ (solid lines) and $S12$ (dashed lines) stress distributions at the contact interface of the sphere contact system in (**b**) and the flat-punch system in (**c**).

However, notable differences remain in the normal and shear stress distributions between the spherical indentation system (Fig. 5b) and the flat-punch system (Fig. 5c), even though they share the same contact radius, as shown in Fig. 5e. In the spherical indentation system, the normal stress is higher and more uniform in the middle part, even though the shear stress is higher at the edge than those in the flat-punch system. These differences are primarily due to the curvature effect in the deformed SMP substrate caused by indentation, which is locked after the R2G transition.

In summary, three mechanisms play key roles in determining the SMP R2G pull-off force as the substrate thickness decreases: (1) increasing the contact radius during press-in, (2) increasing the system stiffness during pull-off, and (3) changing the normal and shear stress distributions due to the curvature of the contact surfaces during pull-off.

## 4.2. Expression of the contact radius

Contact radius for spheres of size $R$=5, 10, 50, and 100 mm, was extracted from the FE simulations, and the $d$-$a$ relations were obtained for different substrate thicknesses. By fitting the $d$-$a$ relations for various thickness-to-contact-radius ratios ($H/a$) into Eqs. (3), the values of the parameters $\alpha = 0.31$ and $\beta = 0.48$ are obtained, and the analytical expressions for predicting the major axis $L$ of the equivalent oblate spheroid and the contact radius $a$ according to the OSE model are given as follows:

$$R/L = 1 - 0.31 \times e^{(-0.48 \times H/a)}, \tag{24}$$

and

$$d = a[1 - 0.31 \times e^{(-0.48 \times H/a)}] \cdot \text{arctanh}\left[(1 - 0.31 \times e^{(-0.48 \times H/a)}) \cdot a / R\right]. \tag{25}$$

As mentioned in Section 2.1, the equivalent oblate spheroid becomes a sphere when $R/L$=1, indicating that the thickness effect of the substrate is negligible. As shown in Fig. 6a, when the thickness-to-contact-radius ratio $H/a$>5, (1−$R/L$) <5%. However, when $H/a$ is less than 5, $R/L$ becomes significantly smaller than 1, indicating that the thickness effect must be considered. Due to the thickness effect of the substrate, a decrease in the relative thickness ($H/a$) leads to an increase in the contact radius. This is equivalent to an increase in the major axis $L$ of the equivalent oblate spheroid under unconfined conditions with an infinitely thick substrate.

To validate the accuracy of the formula in Eq. (25) provided by the OSE model for predicting the contact radius, FE simulations were conducted using a spherical indenter with a radius of $R$=20 mm for different substrate thicknesses. Figure 6b presents the $d$-$a$ relation obtained from the FE simulations (dots) and the theoretical predictions (solid lines) given by Eq. (25). The good agreement between the FE simulations and theoretical predictions confirms the accuracy of the OSE model across a wide range of thicknesses.

Figure 6b also shows that the thickness effect of the substrate leads to an increase in the contact radius for the same indentation depth. This increase in contact radius during press-in is one of the key mechanisms through which the substrate thickness influences the R2G pull-off force.

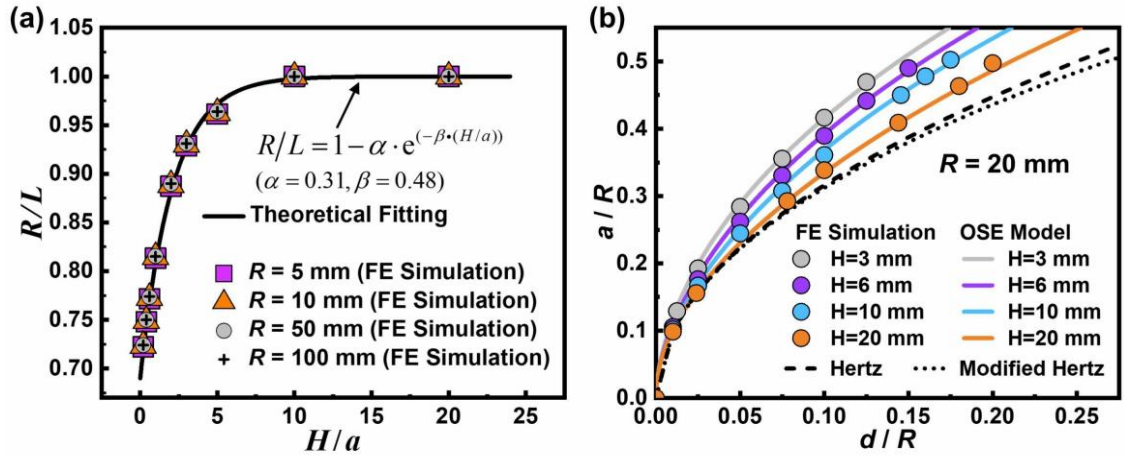

**Fig. 6. Formulation and validation of the contact radius expression based on the oblate spheroid equivalence (OSE) Model using FE simulation results.** (**a**) Fitting of the geometric parameter *R/L* of the equivalent oblate spheroid as a function of *H/a* using the FE simulation results. (**b**) Validation of theoretical predictions from the OSE model (solid lines) against FE simulation results (circles) for a sphere with a radius of 20 mm, not used in model parameter fitting. Predictions for infinite-thickness conditions from the Hertz model (Hertz, 1882) denoted by a dashed line, and the modified Hertz model (Guo et al., 2020) denoted by a dotted line are included for comparison.

In conclusion, this approach demonstrates the equivalence between substrate thickness change in a spherical indentation problem and a change in the shape of the oblate spheroid indenter under infinite-thickness conditions. This leads to the development of a single empirical expression that accurately predicts the contact radius of spherical indenters on finite-thickness substrates, providing a foundation for the subsequent derivation of R2G pull-off forces under confined conditions.

### 4.3. Expression for the pull-off force

To determine the functional forms in Eqs. (21), pull-off forces were extracted from FE simulations using four different indenter radii ($R$=5, 10, 50, 100 mm) on substrates of varying thicknesses, and fitted into Eq. (21). The FE simulation results (dots) indicate that the square of the correction function, $\left[\eta\left(a/H\right)\right]^2$ shows an exponential relationship with the reciprocal of the thickness-to-contact-radius ratio ($a/H$), as depicted in Fig. 7. The fitted expression is given by:

$$\left[\eta\left(a/H\right)\right]^2 = P_c^2/\overline{P}_c^2 = 1 + 28.0 \times \mathrm{e}^{-4.3\times(a/H)}. \tag{26}$$

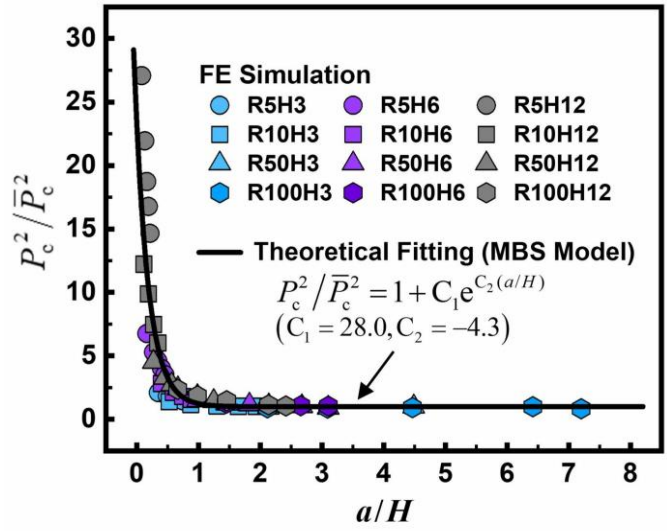

**Fig. 7. Determination of the SMP R2G pull-off force expression under finite-thickness conditions** based on the modified ball-and-socket (MBS) model, fitted to FE simulation results (dots).

The results indicate that the effect of the substrate thickness reaches a plateau (with $\left(1 - P_c^2/\overline{P}_c^2\right) \leq 0.52\%$, less than 1%) when the relative thickness of the substrate is small, i.e., when $a/H > 2$ or $H/a < 0.5$. These cases reduce to the thin-film scenario where the curvature effect is negligible, and the pull-off force can be directly calculated using Eq. (22).

Substituting Eq. (26) into Eq. (21) yields the final expression for the R2G pull-off force according to the MBS model as

$$P_c = \sqrt{(1.0 + 28.0\mathrm{e}^{-4.3(a/H)})\frac{8wk\pi^2 R^6[1-(1-(\frac{a}{R})^2)^{\frac{3}{2}}]^2}{9(R^2-a^2)}}. \tag{27}$$

To verify the accuracy of the MBS model, FE simulations were conducted using a spherical indenter with a radius of 20 mm. The simulated pull-off forces (red square) on substrates of varying thicknesses were compared with the predictions from the MBS model (red solid line) in Fig. 8(b). Additionally, predictions from various FPA models (dashed, dash-dotted, dash-dot-dotted, and dotted lines) are provided for comparison, as shown in Fig. 8.

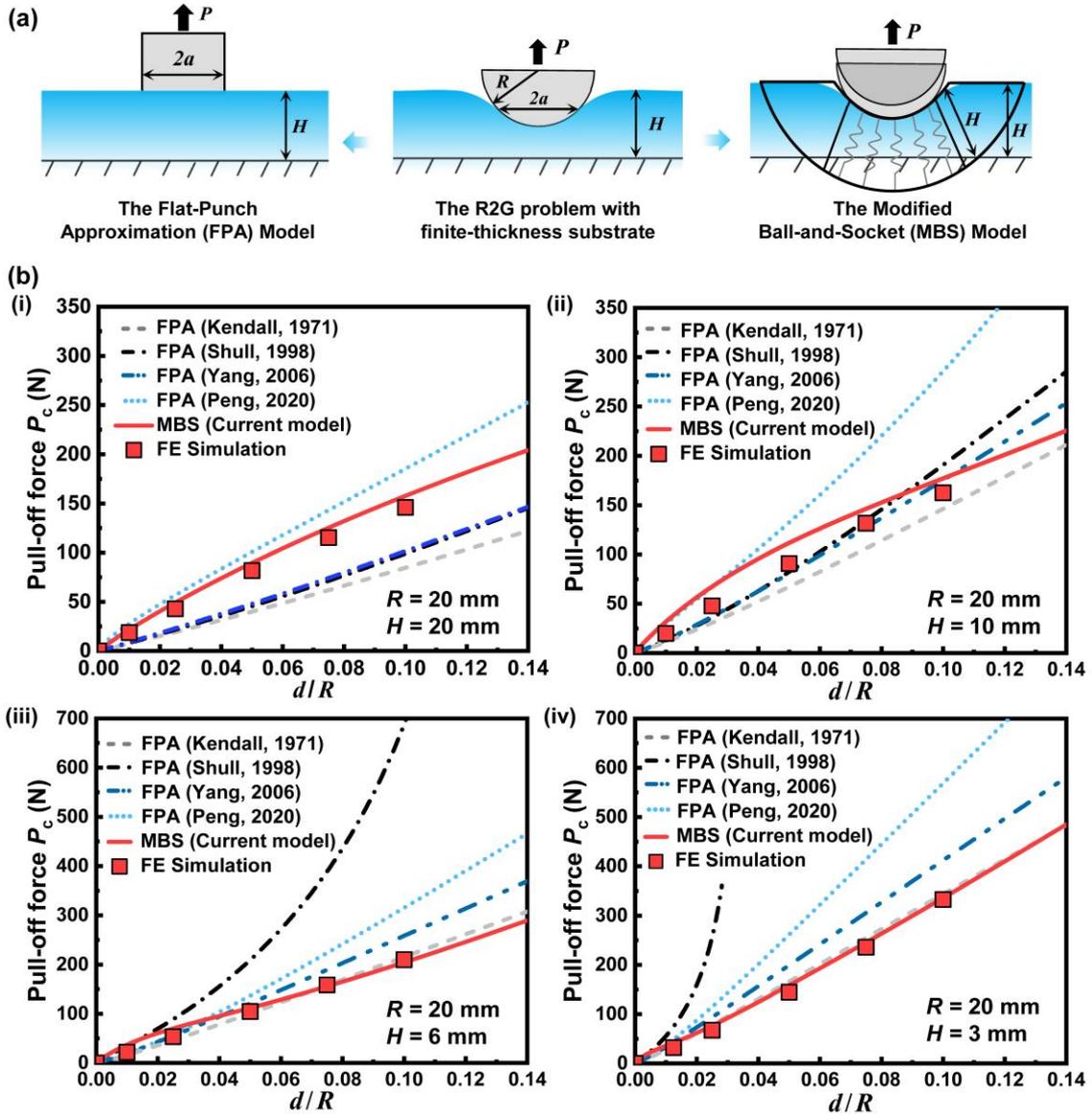

**Fig. 8. Predictions of SMP R2G pull-off forces under finite-thickness conditions.** (**a**) Schematics of the Flat-Punch Approximation (FPA) model and the Modified Ball-and-socket (MBS) model (current model). (**b**) Comparison of FE simulation results (red dots) with theoretical predictions from various FPA models (dashed, dash-dotted, dash-dot-dotted, and dotted lines) and the MBS model (solid lines) under finite-thickness conditions. The spherical indenter radius is $R$= 20 mm, and the SMP substrate thickness $H$ varies from (i) 20 mm, (ii) 10 mm, (iii) 6 mm, to (iv) 3 mm.

The results indicate that the FPA models can only accurately predict pull-off forces under specific conditions. For example, predictions from Kendall's model (dashed lines) agree with FE simulation results when the substrate thickness is small ($H$=3 and 6 mm), but their model significantly underestimates the pull-off forces when the substrate thickness is larger ($H$=10 and 20 mm). None of the FPA models are able to accurately predict the SMP R2G pull-off force across all four thicknesses explored.

In contrast, the proposed MBS model provides accurate predictions of the pull-off force for substrates of a wide range of thickness, as shown by their good agreement with the FE simulation results in Fig. 8(b). The MBS model is further validated by comparison with our previous experimental results (Linghu et al., 2023c), as shown in Fig. 9. The MBS model predictions not only agree well with experimental data when the relative thickness of the SMP adhesive substrate is large ($R$=10 mm, $H$=12 mm, $H/R$=1.2) or small ($R$=10 mm, $H$=50 mm, $H/R$=0.24), but also outperform the FPA model under infinite-thickness conditions by Linghu et al. (2023c), which shows significant deviation from experiments under large indentations in confined conditions ($R$=10 mm, $H$=50 mm, $H/R$=0.24).

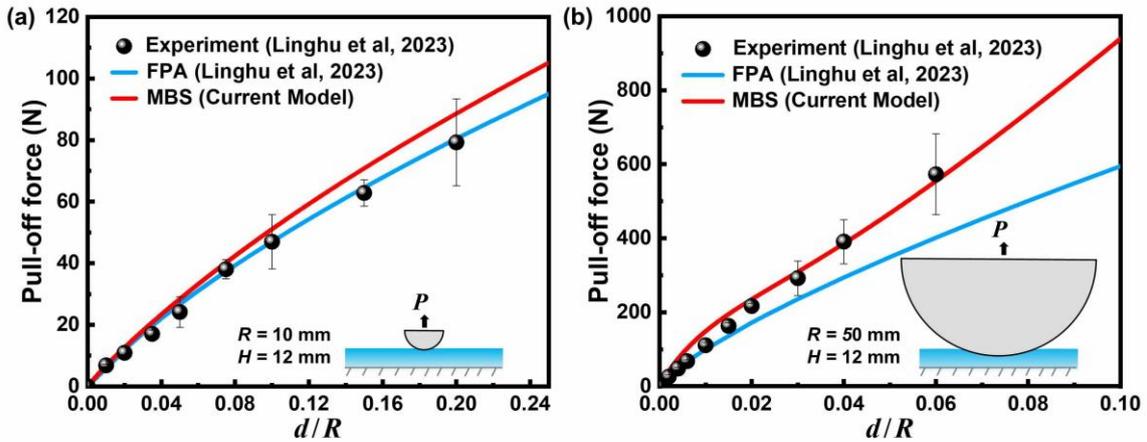

**Fig. 9. Experimental validation of theoretical predictions of SMP R2G pull-off forces under finite-thickness conditions using the modified ball-and-socket (MBS) model.** (**a**) Spherical indenter radius $R$=10 mm and SMP substrate thickness $H$=12 mm. (**b**) Spherical indenter radius $R$=50 mm and SMP substrate thickness $H$=12 mm. Experimental data are sourced from the literature (Linghu et al., 2023c).

## 5. Conclusions

In this paper, we have investigated the finite-thickness effects on SMP R2G adhesion using theoretical modeling and FE simulations, and the model predictions are validated by experimental results from the literature.

Unlike the contact behavior under infinite-thickness conditions, the effect of finite substrate thickness increases the contact radius under the same indentation depth, thus enhancing the SMP R2G pull-off force. Using an oblate spheroid equivalence (OSE) model, an analytical expression for predicting the contact radius under finite-thickness conditions is derived, with the correction parameter determined by fitting the FE simulation results. This model is validated and exhibits excellent agreement with FE simulations conducted for spherical indentation systems with different sphere radii. The OSE model effectively captures the influence of substrate thickness on the contact radius. The results indicate that when the thickness-to-contact-radius ratio ($H/a$) exceeds 5, the effect of substrate thickness on the contact radius becomes negligible. However, when $H/a$ is smaller than 5, the substrate's thickness effect must be considered.

For the pull-off process, the generalization of flat-punch approximation (FPA) models is first discussed under finite-thickness conditions. These models are only applicable for specific thicknesses, as they account solely for the thickness effect while neglecting the curvature effect of the locked deformation in the SMP substrate during press-in due to the shape-locking effect after the R2G transition. To address the R2G adhesion problem under confined conditions, a modified ball-and-socket (MBS) model is proposed to determine the substrate stiffness after shape-locking. The specific form of the pull-off force is derived by fitting FE simulation results. Comparisons with both FE simulation results and experimental data from the literature demonstrate that the proposed MBS model accurately predicts the pull-off force, applicable for substrates of a wide range of thickness. Furthermore, the critical thickness-to-contact-radius ratio, below which the thickness effect cannot be neglected, is identified to be 5.

This work not only uncovers the key mechanisms by which the substrate thickness influences SMP R2G adhesion but also provides accurate analytical expressions for predicting the contact radius and pull-off force. These insights offer valuable design guidance for smart adhesive devices based on these mechanisms in real-world applications.

**CRediT authorship contribution statement**

**Changhong Linghu:** Writing – original draft and revision, Software, Methodology, Investigation, Formal analysis, Data curation. **Wentao Mao:** Writing – original draft and revision, Software, Methodology, Investigation, Formal analysis, Data curation. **Haoyu Jiang:** Formal analysis, Data curation, Validation. **Huajian Gao:** Writing – original draft and revision, Software, Methodology, Investigation, Funding acquisition, Formal analysis, Data curation, Supervision. **K. Jimmy Hsia:** Writing – original draft and revision, Software, Methodology, Investigation, Funding acquisition, Formal analysis, Data curation, Supervision.

**Declaration of Competing Interest**

The authors declare that they have no known competing financial interests or personal relationships that could have appeared to influence the work reported in this paper.

**Data availability**

Data will be made available on request.

**Acknowledgments**

The authors acknowledge the support from the Ministry of Education (MOE) of Singapore under Academic Research Fund Tier 2 (MOE-T2EP50122-0001).

# 6. Appendixes

## 6.1. Appendix A. Existing flat-punch models considering the effect of substrate thickness

There are several models that consider the effect of substrate thickness on the flat-punch problem. The key challenge lies in determining the contact stiffness between the flat punch and a finitely thick substrate. Deriving an analytical expression for contact stiffness for substrates of any thickness is not straightforward. However, when the thickness is much smaller compared to the contact radius, an analytical form of the contact stiffness can be obtained. For example, in the case of a thin film ($H \ll a$), Kendall (Kendall, 1971) provided the stiffness as follows:

$$K_{\text{Kendall}} = \frac{\pi a^2 E}{3(1-2v)H},$$
(A1)

and the pull-off force directly as

$$P_{\text{c, Kendall}} = \pi a^2 \sqrt{\frac{2Ew}{3(1-2v)H}}.$$
(A2)

Another model for thin films was presented by Yang (Yang, 2006). By assuming $H \ll a$, he derived the solution of the stiffness by analyzing the contact problem using elasticity, as follows:

$$K_{\text{Yang}} = \frac{(1-v)E\pi a^2}{(1+v)(1-2v)H}.$$
(A3)

Based on this, the pull-off force was obtained as

$$P_{\text{c, Yang}} = \pi a^2 \sqrt{\frac{2Ew(1-v)}{(1+v)(1-2v)H}}.$$
(A4)

Actually, the model presented by Kendall and Yan is very similar. They are related by

$$P_{Yang} = P_{Kendall} \sqrt{\frac{3(1-v)}{1+v}}.$$
(A5)

While deriving an analytical expression for stiffness across various thicknesses is challenging, several empirical formulas have been proposed for finite-thickness conditions.

Shull et. al. (Shull et al., 1998) analyzed the contact problem using FE simulations across different thicknesses and got an empirical formula for the stiffness as follows:

$$K_{\text{Shull}} = \frac{2Ea}{1-v^2}\left[1+\left(\frac{0.75}{\left(\left(a/h\right)+\left(a/h\right)^3\right)}+\frac{2.8\left(1-2v\right)}{\left(a/h\right)}\right)^{-1}\right]. \qquad \text{(A6)}$$

Based on this, the pull-off force was obtained as

$$P_{\text{c, Shull}} = \sqrt{\frac{4\pi RawK_{\text{Shull}}^{2}}{\partial K_{\text{Shull}}/\partial a}} \qquad \text{(A7)}$$

Similarly, Hayers et. al. (Hayes et al., 1972) utilized a correction function $k(H/a, v)$ to consider the stiffness of the infinite-thickness substrate as follows:

$$K_{\text{Peng}} = \frac{2Ea}{1-v^2}k(H/a, v). \qquad \text{(A8)}$$

Based on this, Peng et. al (Peng et al., 2020) got the pull-off force

$$P_{\text{c, Peng}} = \sqrt{\frac{8\pi wEa^3}{1-v^2}\psi\left(\frac{H}{a}, v\right)}, \qquad \text{(A9)}$$

where $\psi\left(\dfrac{H}{a}, v\right)$ is a function of the thickness-to-contact-radius ratio $H/a$ and poison ratio $v$, the exact form of $\psi\left(\dfrac{H}{a}, v\right)$ is very complex, so it is not given here. Readers who are interested can read the reference (Peng et al., 2020) for more details.